\documentclass[utf8]{FrontiersinHarvard} 

\usepackage{url,hyperref,lineno,microtype,subcaption}
\usepackage[onehalfspacing]{setspace}

\def\keyFont{\fontsize{8}{11}\helveticabold }
\def\firstAuthorLast{Emig {et~al.}} 
\def\Authors{Thorsten Emig\,$^{1,*}$, Guillaume Adam\,$^{2}$}
\def\Address{$^{1}$Laboratoire de Physique
Th\'eorique et Mod\`eles Statistiques, CNRS UMR 8626,
Universit\'e Paris-Saclay, 91405 Orsay cedex, France, 
$^{2}$Coaching 4.0 SAS, 166 Chemins des Varons, 73370 Le-Bourget-du-Lac, France}
\def\corrAuthor{Thorsten Emig}

\def\corrEmail{thorsten.emig@cnrs.fr}

\begin{document}
\onecolumn
\firstpage{1}

\title[Physiology of World Running Records]{Evolution of World Running Record Performances for Men and Women: Physiological Characteristics} 

\author[\firstAuthorLast ]{\Authors} 
\address{} 
\correspondance{} 

\extraAuth{}

\maketitle

\begin{abstract}
Running world records (WRs) contain information about physiological characteristics that determine running performance. The progression of WRs over time encode the evolution of these characteristics.
Here we demonstrate that a previously established model for running performance describes WRs since 1918 for men and since 1984 for women with high accuracy. 
The physiological parameters extracted from WR for each year are interpreted in terms of historical changes in training approaches and corresponding physiological adaptions, technological progress, social effects, and also the use of performance enhancing drugs. While the last two decades had witnessed stagnation of WRs, recent improvements in endurance have enable new WRs, presumably aided by recent technological advancements.
\tiny
 \keyFont{ \section{Keywords:} running, performance, physiology, world records, mathematical model} 
\end{abstract}

\section{Introduction}

Scientists have been fascinated by trying to explain running performance and to predict its limitations for more than 100 years. Over time, new training methodologies, an increasing number of athletes, and hence attempts to break records, and also the use of performance-enhancing drugs (doping) have contributed to a substantial improvement of human running performance. Endurance running is a preferred sport to measure accurately required physiological profiles as external factors are not so much subject to variations as in cycling, due to a similar running track shape and road race courses that have experienced little changes over recent decades.
Important for the density of new running records could be the number of competitive opportunities per year.
Other factors could be that humans are reaching the biological limits for performance and the potential role of technical innovations in training and equipment gains importance in more recent times.
With the goal to prevent the apparently wide spread doping in elite sports competition, the World Anti-Doping Agency (WADA) was created in 1999 \citep{Kruse:2014wi} and developed and applied drug testing protocols. 
If historical improvements in athletic performance have  benefited  from doping, then improved doping control should be reflected by a saturating or even a declining performance.
An important mark in doping control is the introduction of techniques to detect the use of exogenous erythropoietin (EPO). A first test for EPO was introduced at the 2000 Summer Olympic Games. 

The aim of this work is twofold. First, we intend to demonstrate that running world records (WRs) of more than a century can be described accurately by a simple, previously introduced mathematical model with four performance parameters. These parameters quantify different physiological characteristics of a runner or  of a group of runners. Second, we would like to relate  changes of the performance parameters over time to historical events such as changing training methodologies, social effects and the use and the control of performance enhancing drugs.
To probe the predictive power of our model, we also compute the expected times for yearly world records for half marathon and marathon from the world records on shorter distances of the corresponding year.
We note that  half marathon and  marathon are run on roads, over varying terrain with different surface properties, and with some small uncertainty in actually run distance since an athlete might not always choose the shortest trajectory between start and finish as assumed in official course measurement. Other factors result from exposure to external performance limitations due to climate, in particular heat and wind.
All considered WRs on shorter distances were set on a 400m long running track, implying very accurate knowledge of actually run distance, optimal running surface, and more controlled wind conditions.

One might wonder if the observed WRs could be explained solely by an increased number of attempts, i.e., by an increasing number of competitive athletes and/or number competitions. However, it can be estimated that over
the last 100 years, the observation of new world records has been too frequent to be explained solely by {\it random fluctuations} \citep{Adam:2016aa}. This is to say that records exceed the mean number of records that would be observed after a certain number of attempts by athletes which constitute a probability distribution of finishing times that does not change over time due to the absence of {\it systematic improvement}. In particular the half and full marathon has witnessed substantial performance improvements in recent years. Interestingly, the finishing times are  correctly predicted  by our model based on previously achieved, stable physiological performance parameters estimated from records on shorter distances. 
However, the physiological performance parameters deduced from men world records on distances up to 10.000m show saturation over the last two decades. 
This saturation suggest that the recent improvements of half and full marathon records is due to a delayed realization of a "hidden" physiological potential. In fact, our model suggests that since about 1995 half and full marathon times had been too slow compared to times on shorter distances, see Fig.~\ref{fig:1}(c).  
One reason for this uncovering of physiological potential could be the increasing number of east African athletes participating in half and full marathons, as witnessed by the African dominance in record breaking athletes over the last two decades. Another reason is likely the emergence of new running shoe technology 
since 2016 which is now used by most leading elite athletes competing in road races, typically over distances of half and full marathon \citep{Muniz_Pardos_2021}. Similar developments in  track spike shoe technology seem to be responsible for recent improvement of records for 5.000m and 10.000m, both for women and men, which come after at least a decade of no new records.

When modern Olympic Games started in 1896, women had to wait until 1928 to have the right to participate in athletics events. The only middle distance allowed was the 800m but after the event took place, the IOC decided to ban this event for women until 1960. Women got the right to participate in events longer than 1.500m only at the Olympic Games in 1984, including the marathon. Hence, 
early running performances of women, and also for the half marathon in general, were subject to strong fluctuations. Therefore we restricted our analysis for women to the time after 1983
since women were not routinely permitted to participate in long distance racing until the 1970th and performances of women improved substantially in the early years, presumably due to the strong increase in racing opportunities.  At about the same time, around 1984, half marathon performances started to become more consistent with other long distance performances, both for men and women. Therefore, we limited the inclusion of half marathon world records to the time after 1984 for men.

\section{Materials and Methods}

\subsection{Mathematical Model}

Various mathematical models for running performance have been proposed and applied to data for running records; for a recent review see \citep{Sreedhara:2019ud}. We had derived a minimal mathematical model from basic physiological principles which reproduces the important real world observation that the mean running velocity declines only logarithmically with duration \citep{Mulligan:2018ue}. Our model has been successfully applied to a large data set of running activities of non-professional athletes to predict their performance on the marathon distance \citep{Emig:2020wj}. In the following we briefly describe the rational behind our model and provide the main equations. For further details we refer to \citep{Mulligan:2018ue}.

The model connects running velocity $v$ and the metabolic power $P(v)$ that is required to run at the velocity $v$ by the linear relation
\begin{equation}
  \label{eq:economy}
  p(v) = \frac{P(v)-P_b}{P_m-P_b} = \frac{v}{v_m} \,.
\end{equation}
Here $P_m$ corresponds to maximal aerobic power (MAP) associated with maximal oxygen uptake VO$_{2max}$, $v_m$ is a crossover velocity which is the smallest velocity that elicits MAP, so that it is expected to correspond to the maximal aerobic speed (MAS). More formally, MAS is defined as the smallest speed at VO$_{2max}$ during a test protocol with an incrementally increasing speed. During longer efforts, due to a drift in oxygen consumption, MAS might not always correspond exactly to  MAP.
 In above equation $P_b$ is the basal metabolic power. More precisely,  $P_b$ is associated with the power that is obtained by linearly extrapolating the linear branch of the power--velocity relation to zero velocity, neglecting non-linear behavior for sub-running (walking) velocities \citep{morgan1989factors}. The relation of Eq.~\ref{eq:economy} is expressed relative to the power reserve $P_m-P_b$ which shows that fractional power reserve $p(v)$ is given by the velocity measured in units of $v_m$. This linear relation is also known as running economy. The velocity $v_m$ measures the combined effect of running economy and MAP.
For running intensities that do not require more power than provided aerobically, $p$ varies between $0$ and $1$. 
In the following, we denote the time over which MAP and hence a velocity $v_m$ can be maintained by $t_c$.
For velocities higher than $v_m$, anaerobic power supply leads to a fractional power reserve $p>1$ which can be maintained only for a duration shorter than $t_c$. We shall show below that our model depends only on the fractional power reserve. Hence no information on the power reserve $P_m - P_b$ is required.

The second important concept is that of endurance. It can be defined from the relation $P_\text{max}(T)$ between the maximal possible average the power output $P_\text{max}$ during a competition (or exercise session) and its duration $T$.  Naturally, $P_\text{max}(T)$ decreases with $T$. First models suggested a power law decay, like Hill's "running curve" from 1925, which stated $P_\text{max}(T)=E_n/T+P_a$ with anaerobic energy $E_n$ and aerobic power $P_a$ \citep{Hill:1925pd}. However, it is known that aerobic power cannot maintained for an infinite duration. Hence, the "running curve" appears incomplete for longer times $T$. A number of modifications of the curve have been postulated in the literature  \citep{Sreedhara:2019ud}. A derivation from basic physiological principles has been presented recently \citep{Mulligan:2018ue}, and we shall reformulate the approach here. 
Consider the following {\it Gedankenexperiment}. Let us assume that a runner races a certain distance in a time $T$, following two different pacing strategies: (Strategy 1) The runner follows what one would probably call a realistic strategy, namely run the entire distance with a constant power output $P_\text{max}(T)$, i.e., with maximally sustainable power over time $T$. (Strategy 2) Assume (hypothetically) that the runner can overcome fatigue and increase power output over time such that at each moment $t$ of the race the used power is the maximally sustainable power $P_\text{max}(T-t)$ which the runner can sustain for the remaining time $T-t$ of the race. Of course, in (Strategy 2) additional energy is required to account for compensating fatigue. This supplemental energy should increase linearly with time since factors that lead to fatigue increase typically with a constant rate. Due to different physiological processes for energy generation during anaerobic and aerobic exercise, the rate is not constant but can show a crossover behavior at the time $t_c$ which sets the limit for the duration over which power greater than MAP can be sustained. When we denote the supplementary energy by $E_\text{sup}(T)$, it can be parametrized as follows (with rates $\gamma_{an}$, $\gamma_{ae}$):
\begin{equation}
  \arraycolsep=1.4pt\def\arraystretch{1.8}
  E_\text{sup}(T) = \left\{
    \begin{array}{ll}
    \gamma_{an} T (P_m - P_b)   
  & \text{ for } T \le t_c \\[1em]
    \left[ \gamma_{an} t_c + \gamma_{ae} (T-t_c)\right] (P_m - P_b)   
      & \text{ for } T > t_c
      \end{array}
 \right. \, ,
\end{equation}
where we used the power reserve $P_m - P_b$ as a natural scale to measure power, and assumed a sharp crossover. 
When the energy $E_\text{sup}(T)$ exactly compensates fatigue as required for (Strategy 2), a consequence of our Gedankenexperiment is that energy balance implies
\begin{equation}
\label{eq:balance}
T \, P_\text{max}(T) + E_\text{sup}(T) = \int_0^T P_\text{max}(T-t) dt \, .
\end{equation}
This simple integral equation for $P_\text{max}(T)$ has to be solved with the condition that at time $T=t_c$ the maximally sustainable power is MAP $P_m$ since this is how we had defined the time $t_c$ before.  Hence we require $P_\text{max}(t_c)=P_m$. The solution is 
\begin{equation}
  \label{eq:Pmax_solution}
  \arraycolsep=1.4pt\def\arraystretch{1.8}
  P_\text{max}(T) = \left\{
    \begin{array}{ll}
    P_m - \gamma_{an} (P_m - P_b) \log  \frac{T}{t_c} 
  & \text{ for } T \le t_c \\[1em]
 P_m - \gamma_{ae}  (P_m - P_b) \log  \frac{T}{t_c} 
      & \text{ for } T > t_c
      \end{array}
 \right. \, .
\end{equation}
Using Eq.~(\ref{eq:economy}) with $P(v=v_\text{max})=P_\text{max}(T)$, this maximally sustainable power can be converted into a maximally sustainable velocity $v_\text{max}(T)$ that can be sustained for a duration $T$,
\begin{equation}
  \label{eq:vmax_solution}
  \arraycolsep=1.4pt\def\arraystretch{1.8}
  v_\text{max}(T) = \left\{
    \begin{array}{ll}
    v_m \left( 1 - \gamma_{an} \log  \frac{T}{t_c} \right)
  & \text{ for } T \le t_c \\[1em]
 v_m \left( 1 - \gamma_{ae} \log  \frac{T}{t_c} \right)
      & \text{ for } T > t_c
      \end{array}
 \right. \, .
\end{equation}
By setting $v_\text{max}=d/T$ and solving for $T$
one obtains the shortest possible time $T(d)$ for covering a distance $d$.
The solution
can be expressed as the real branch $W_{-1}(z)$ of the Lambert
W-function which is defined as the (multivalued) inverse of the
function $w\to w e^w$ \citep{Corless:1996fh},
\begin{equation}
  \label{eq:T_of_d}
  \arraycolsep=1.4pt\def\arraystretch{1.8}
  T(d) = \left\{ 
    \begin{array}{ll}
       - t_c \frac{d}{\gamma_\text{an} d_c} / W_{-1}\left[-\frac{d}{\gamma_\text{an} d_c }
      e^{-1/\gamma_\text{an}}\right] \, & \text{ for } \, d \le d_c\\
      - t_c \frac{d}{\gamma_\text{ae} d_c} / W_{-1}\left[-\frac{d}{\gamma_\text{ae} d_c }
      e^{-1/\gamma_\text{ae}}\right] \, & \text{ for } \, d \ge d_c
      \end{array}\right. \, ,
\end{equation}
where we have defined the  distance $d_c=v_m t_c$. [The
  function $W_{-1}(z)$ is real valued for $-1/e \le z <0$, a condition
  which is fulfilled for all distances $d$ that we consider.] Note that $T(d)$
is continuous at $d=d_c$ with $T(d_c)=t_c$ since $W_{-1}(w
e^w)=w$ holds. Now that we have derived all main equations of our model, we can compare them to running WRs.

\subsection{Data: Running World Records}

Running WRs for both men and women were obtained from the 
Statistics Handbook of the IAAF (Doha 2019) \citep{IAAF}, and most recent times from online sources. \footnote{Wikipedia, Index of athletics record progressions, \url{https://en.wikipedia.org/wiki/Index_of_athletics_record_progressions}} The IAAF handbook lists the progressive WR performances, as ratified by the IAAF or the F\'ed\'ederation Sportive F\'eminine Internationale (FSFI), which was absorbed by the IAAF in 1936. Covered in the listings are all the WRs currently listed under IAAF Rule 261 in 2018/2019 IAAF Competitions. For distances different from the half marathon and marathon, the WRs were obtained on a 400m athletic track.  For details on official acceptance and rules, see the Statistics Handbook of the IAAF \citep{IAAF}.

\subsection{Computation of Model Parameters}

Our mathematical model depends on the four independent parameters $v_m$, $t_c$,
$\gamma_\text{an}$ and $\gamma_\text{ae}$. In the present work, a set of these four parameters characterizes a group of runners. Here the group consists of the male or female
world record holders in a given year.  Otherwise the
model is universal in the sense that it does not contain any additional fixed
parameters or constants. The four parameters can be estimated from a
given set of performance results (average velocity and time for a given set of distances) from exercise performed at relative maximal intensity, i.e., from WR performances. To compute the model
parameters, we have fitted by the least squares method the continuous piecewise linear function of Eq.~(\ref{eq:vmax_solution}) to the data pairs $(v,\log T)$ for all WRs set on the track (1000m to 10.000m), using the approach outlined in \citep{Kundu}. We note that the breakpoint at $t_c$ between the linear segments is itself a variable, making the regression problem more challenging.

\section{Results}

\subsection{Evolution of Running World Records}

\begin{figure}[h!]
\begin{center}
\includegraphics[height=0.9\textheight]{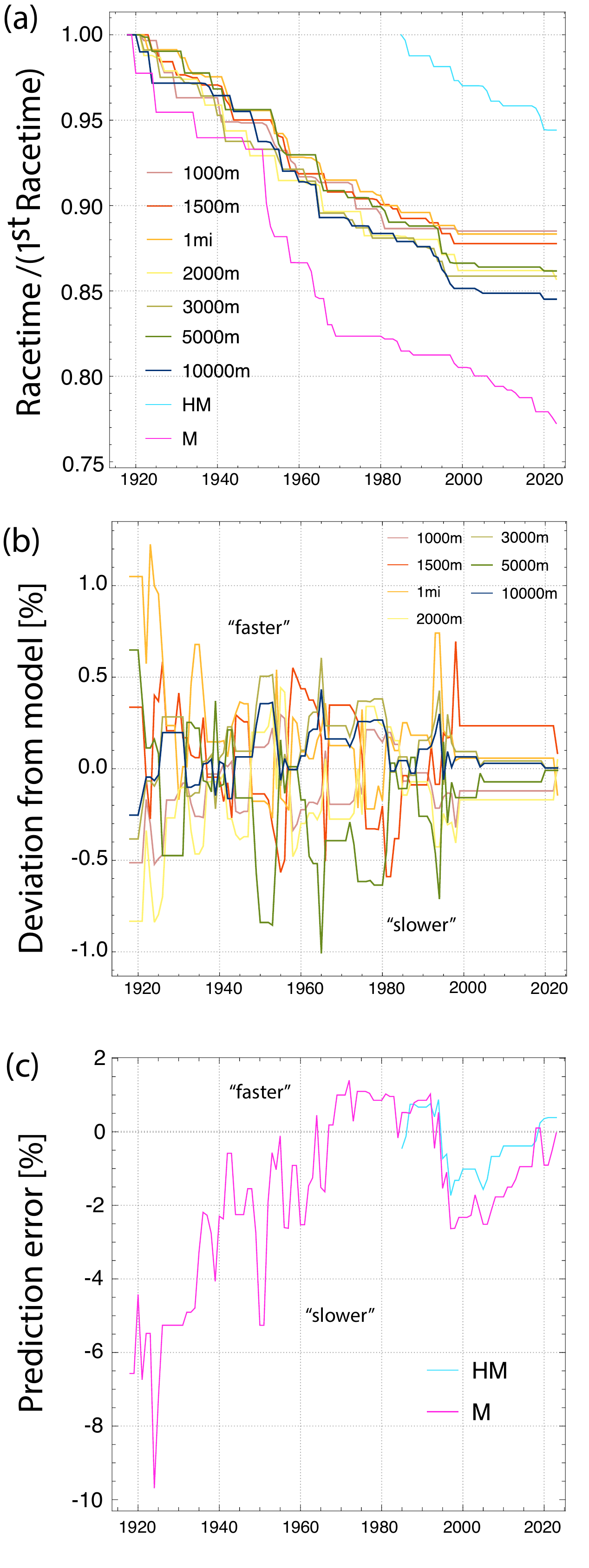}
\end{center}
\caption{{\bf Men World Records from 1918 to 2023: (a) Evolution of record times relative to the records in 1918 (for half marathon first result for 1985), (b) Relative deviation between time from fit to model and actual record times (positive/negative deviation means actual record is faster/slower than model fit), (c) Relative deviation of model prediction from actual record times for marathon (M) and half marathon (HM).}}
\label{fig:1}
\end{figure}

\begin{figure}[h!]
\begin{center}
\includegraphics[height=0.9\textheight]{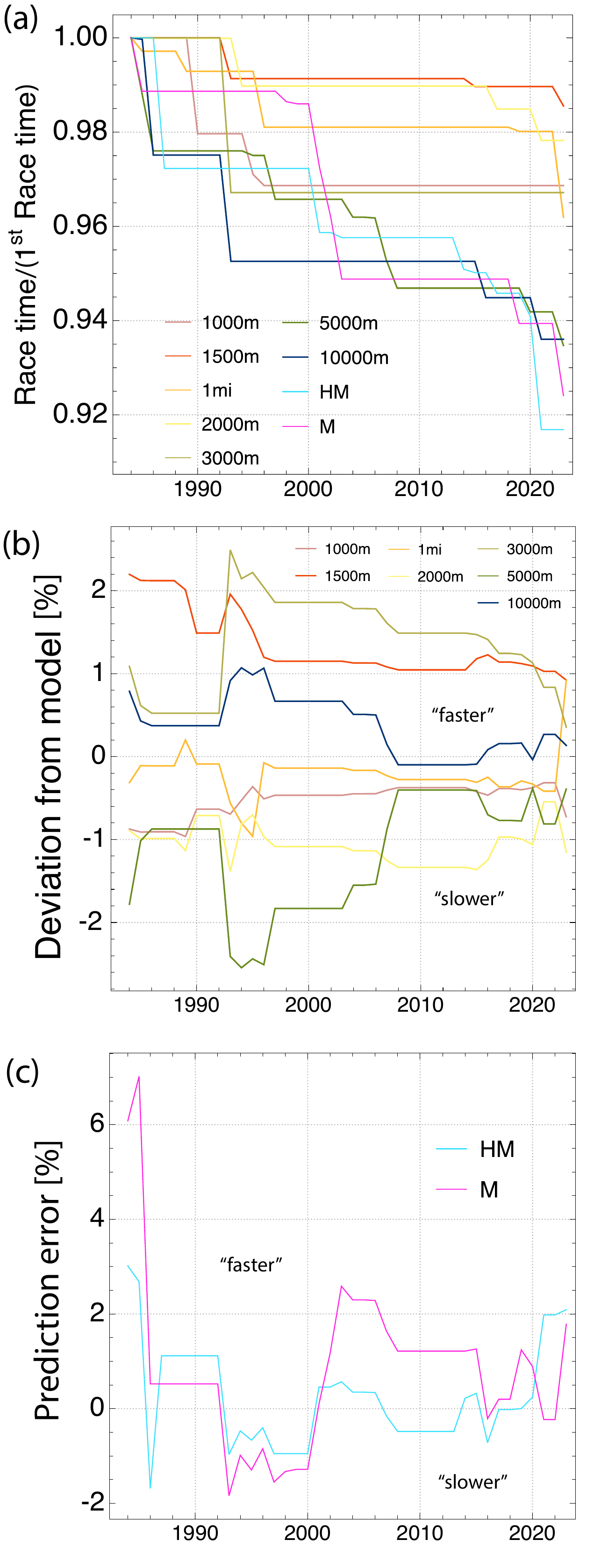}
\end{center}
\caption{{\bf Women World Records from 1984 to 2023: (a) Evolution of record times relative to the records in 1984, (b) Relative deviation between time from fit to model and actual record times (positive/negative deviation means actual record is faster/slower than model fit), (c) Relative deviation of model prediction from actual record times for marathon (M) and half marathon (HM).}}
  \label{fig:2}
\end{figure}

The evolution of the time series of running world records is depicted in Fig.\ref{fig:1}(a) for men (since 1918) and in Fig.\ref{fig:2}(a) for women (since 1984). To better visualize the change of WR times, the graphs show them relative to the time in the first year of the series.  There are interesting similarities and differences between the evolution for men and women. For men, the Marathon WR followed the general trend of other distances until about 1950 after which it improved much faster until today. For women, while the time series is much shorter, the improvement of the Marathon and Halfmarathon performances follows closely the evolution of WRs for 5.000m and 10.000m, until today. For men and women, the WR times for track middle distances from 1.000m to 3.000m have improved continuously until about 1995 after which they stagnated until more recently (for women) or even until today (for men, with the exception of a recent WR for 2.000m). For men, the improvements for 5.000m and 10.000m also decreased substantially after about 1995. 

We have computed the four parameters of our model from the WRs for each year, as outlined before, and then used Eq.~(\ref{eq:T_of_d}) to determine the time predicted by our model for a given track distance between 1.000m and 10.000m. The deviation between the actual and predicted time provides a measure for the overall consistency of the WR performances with our model. It is shown in percent in Fig.~\ref{fig:1}(b) for men and in Fig.~\ref{fig:2}(b) for women. For men, the WRs for each year appear to be mutually consistent and in very good agreement with our model, with the deviation typicaly bounded between $\pm 0.5\%$. The consistency is less good for women, with a typical deviation of less than $\pm 2\%$. Interestingly, over more recent years, both for men and women, times have become increasingly consistent, and agreement with our model has increased overall. In 2023 the average agreement is $0.04\%$ for men and $0.66\%$ for women (see also Figs.~\ref{fig:3}(a) and \ref{fig:4}(a)).

\subsection{Model Parameters for World Records}

Real world data like running world record performances contain very useful information about maximized physiological response, and can be used to validate theoretical models that have been derived entirely from bio-energetic considerations.
An advantage over laboratory experiments and tests is a more realistic and competitive environment during the running events, with presumably maximal motivation of the athlete and optimal preparation. Clearly, there are also disadvantages of real world data, including less controlled climate and environmental conditions, varying characteristics of the racing course (for Marathon and Halfmarathon), and the lack of the possibility to measure physiological variables such as heart rate, blood lactate concentration, and oxygen consumption, to name a few. Nevertheless, one might be able to extract physiologically meaningful information from best times on a certain set of distances which characterizes either an individual athlete or a group of athletes in the case of WRs. In the latter case, considered here, the physiological model parameters describe the "optimal" athlete who has performed the best possible training for every given distance. At every given year, the physiological parameters are like a fingerprint of the best possible human running performance at that time.

A mathematical model which should be able to describe of group of athletes must be independent of any {\it fixed} athlete specific quantities such as body weight, running economy (power output required to run at a given speed), peak oxygen uptake, and others \citep{Peronnet:1989dp}. Our model fulfills this
requirement, and in this section we shall validate its accuracy by comparing it to WR performances on distances from 1000m to 10.000m. When estimating model parameters, we do not consider marathon and halfmarathon since both are road races and hence course and environmental conditions are less controlled than for a running track, giving rise to fluctuations that are not intrinsic to human physiology. 

WRs and other running records have been analyzed before and found to follow
an approximate power law, i.e., $v_\text{max} \sim T^{-\beta}$ \citep{Kennely:1906aa,Riegel:1981ci,Savaglio:2000wi}. 
However, the exponent $\beta$ of this power law
shows variations with distance which renders its
universality and general applicability questionable.  Also there is
no physiological foundation for a simple power law. In fact, the
existence of a crossover velocity $v_m$ implies different scaling of
performances below and above this velocity due to distinct
physiological and bio-energetic processes involved. A model with a broken (piecewise) power law had been proposed  \citep{FREDERIK:1959uc}. However, the breakpoint appears at too short times, presumably since distances below 1000m had been included in the fit. As basic principles of energy balance underlying our model predict a different scaling of the form $v_\text{max} \sim \log T$, it is important to compare this prediction to WR performances. 
Following the method described in the previous section, we have estimated the
parameters of our model for each year from the men and women WRs.  

\begin{figure}[h!]
\begin{center}
\includegraphics[height=0.9\textheight]{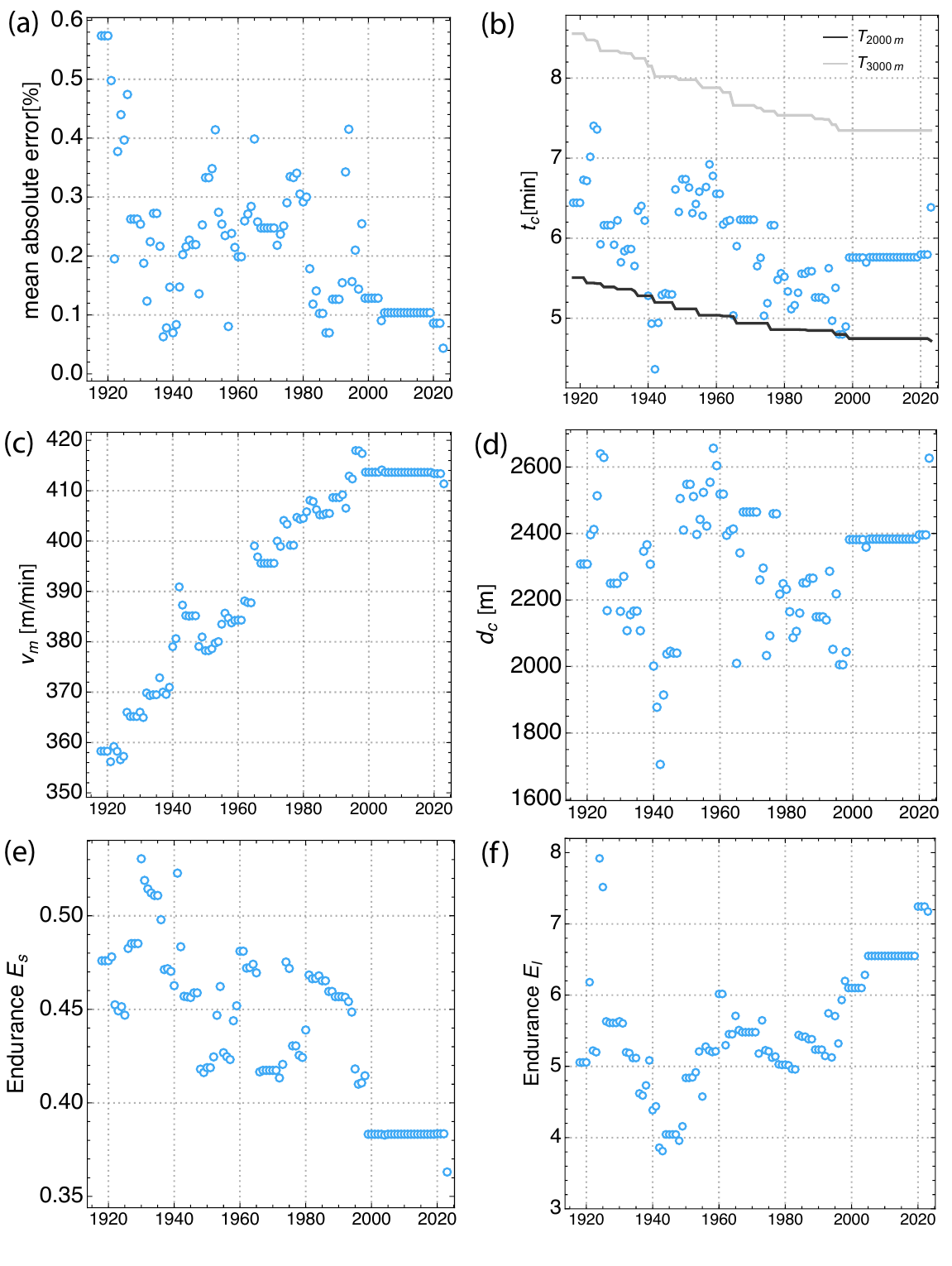}
\end{center}
\caption{{\bf Evolution of model parameters for Men World Records from 1000m to 10.000m, for 1918 to 2023: (a) Mean of absolute deviation between model fit and actual records, (b) crossover time $t_c$ and race times $T_{2000m}$ and $T_{3000m}$ for 2000m and 3000m, respectively, (c) maximal aerobic velocity $v_m$, (d) crossover distance $d_c$ (e) stort duration anaerobic endurance $E_\text{s}$, (f) long duration aerobic endurance $E_\text{l}$.}}
  \label{fig:3}
\end{figure}

\begin{figure}[h!]
\begin{center}
\includegraphics[height=0.9\textheight]{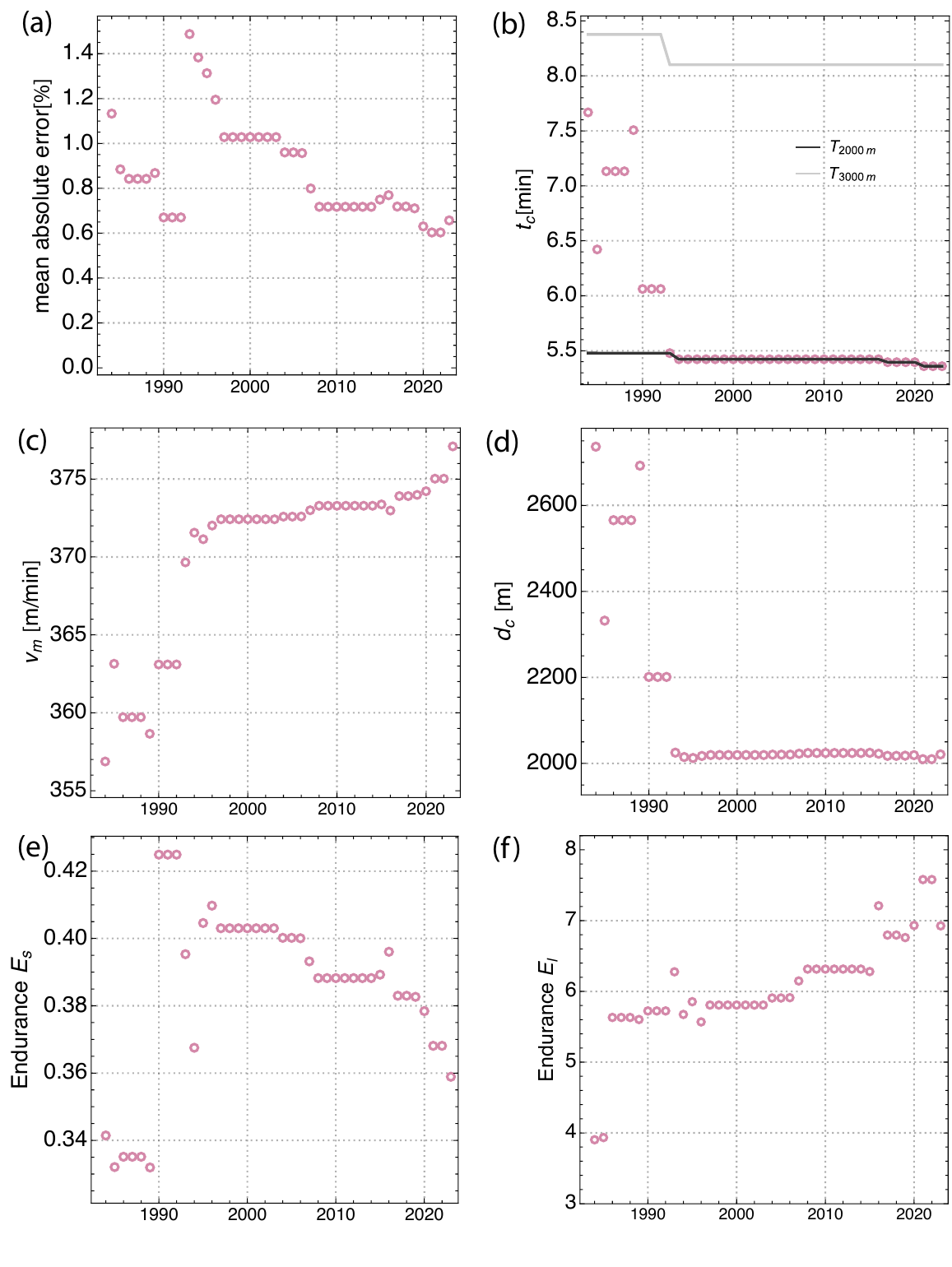}
\end{center}
\caption{{\bf Evolution of model parameters for Women World Records from 1000m to 10.000m, for 1984 to 2023: (a) Mean of absolute deviation between model fit and actual records, (b) crossover time $t_c$ and race times $T_{2000m}$ and $T_{3000m}$ for 2000m and 3000m, respectively, (c) maximal aerobic velocity $v_m$, (d) crossover distance $d_c$ (e) short duration anaerobic endurance $E_\text{s}$, (f) long duration aerobic endurance $E_\text{l}$.}}
  \label{fig:4}
\end{figure}

The parameters $\gamma_\text{an}$ and $\gamma_\text{ae}$ can be viewed as characterizing endurance.
Therefore we define an endurance for the dominantly {\it aerobic} or long distance range as
$E_\text{l}=\exp(0.1 /\gamma_\text{ae}) > 1$ so that the duration over which a runner can
maintain $90\%$ of MAS is given by
$T_{90\%}=t_c E_\text{l} > t_c$. Hence a smaller $\gamma_\text{ae}$, and a larger $E_\text{l}$, corresponds to better
endurance. Similarly, an endurance for the dominantly {\it anaerobic} or short distance range can be defined as $E_\text{s}=\exp(-0.1 /\gamma_\text{an}) < 1$ so that a runner can sustain $110\%$ of MAS for a duration of $T_{110\%}=t_c E_\text{s} < t_c$. Opposite to the aerobic  range, here a larger $\gamma_\text{an}$ corresponds to a better endurance, and hence a larger $E_\text{s}$. The choice of $90\%$
and $110\%$ of MAP is arbitrary, and other sub- and
supra-maximal values could be chosen to define endurance without any
qualitative difference in interpretation. 
The resulting parameters
$t_c$, $v_m$ together with the endurances
$E_\text{s}$ and $E_\text{l}$ are shown in Fig.~\ref{fig:3} for men and in Fig.~\ref{fig:4} for women.
The figures also show the mean absolute model error, averaged over all distances between 1000m and 10.000m, and the crossover distance $d_c=v_m t_c$. The exact yearly WR times and predicted times for all distances, together with the model parameters and endurances are provided in the Tables in the Suppelemental Material. 

In the following we describe the evolution of model parameters in relation to the evolution of WR times. An interpretation of this evolution in the context of training methodology and other factors is provided in the next section. An important result is the excellent agreement between our model and the WR times for distances between 1.000m and 10.000m. For men, the model error has been fluctuating between $0.1\%$ and $0.4\%$ and then plateaued at $0.1\%$ since 2005. Only very recently a further reduction to $0.04\%$ could be observed, see Fig.~\ref{fig:3}(a).
For women, the model error has been consistently decreasing from about $1.5\%$ to about $0.6\%$ nowadays, see Fig.~\ref{fig:4}(a). This observation means that our physiological model parameters are sufficient to describe the WR for all distances with high precision. Hence, the time evolution of these parameters encodes the  improvement of relevant physiological characteristics of world class runners over the last decades. 

The aerobic power has increased in a {\it linear} fashion over the years, both for men and women. This can be seen from the increase of the MAS $v_m$, see Figs.~\ref{fig:3}(c), \ref{fig:4}(c). While for women the increase seems unbroken, for men the linear increase has stopped suddenly around the year 2000 and plateaued afterwards. 
When considering MAS, it is also important to take into account the 
crossover time $t_c$ which determines for how long a runner can sustain the MAS.
We observe that $t_c$ is decreasing, following approximately the mean of the WR time for 2.000m and 3.000m competitions, see Figs.~\ref{fig:3}(b), \ref{fig:4}(b). 
For men, over the last decades the crossover time has approached a value which is  slightly less than 6 minutes. For women, the value is found to be about 5.5 minutes. Interestingly, a series of different laboratory measurements on runners has found a time of $t_c = (5.92 \pm 1.02)$min \citep{Bosquet:2002aa}. However, a word of caution is in order here: Since there exist no WR times for distances between 2.000m and 3.000m, a more precise resolution of $t_c$ from the available data is not possible. Our model provides strong evidence that $t_c$ is located between the times for 2.000m and 3.000m, but for a more precise estimate of $t_c$ best performances for intermediate distances would be required as an input to our model. The maximal aerobic distance $d_c$, which measures the combined effect of MAS $v_m$ and the time $t_c$, shows no obvious trend but fluctuates around a mean value which is located between the 2.000m and 3.000m race distances, both for men and women, see Figs.~\ref{fig:3}(d), \ref{fig:4}(d). For both men and women, no very clear trend of short distance endurance $E_\text{s}$ while for men there seems to be a minor decrease from about $0.5$ to about $0.4$, Figs.~\ref{fig:3}(e), \ref{fig:4}(e). Contrary, the long distance endurance $E_\text{l}$ has been increasing both for men and women from minimal values of about $4$ to more than $7$ in recent times, see Figs.~\ref{fig:3}(f), \ref{fig:4}(f).

\subsection{Prediction of Times for Half and Full Marathon}

The knowledge of the physiological parameters for each year allows to compute predicted finishing times for half and full marathon WRs, using Eq.~(\ref{eq:T_of_d}). 
Comparing the model predictions with the actual WR times on these two long distances provides another important test of our model and the employed concept of endurance. Note that for these long distances, only the long distance endurance parameter $E_\text{l}$ is relevant. 
The prediction error (relative difference between predicted and actual WR time) for Marathon and Halfmarathon is shown in Fig.~\ref{fig:1}(c) for men and in Fig.~\ref{fig:2}(c) for women. This error is not necessarily due to a potential misconception of the model. Instead, it can also arise from an intrinsic inconsistency between the performances on shorter distances and those on the half and full marathon distances. Non-optimal training might have prevented athletes from reaching the physiologically maximal possible potential. In fact, the time evolution of the error for the men's WR points to the latter interpretation. While the Marathon WRs have been much too slow between 1918 and 1965, they have become consistent with shorter distances ($\simeq \pm 2\%$) in recent decades. 
After substantial improvements of WR's for 5.000m and 10.000m in the 1990th, it took almost two decades for the marathon and halfmarathon times to catch up. Since a few years they display very good consistency with all shorter distances. For women, while the time series of WR is much shorter than for men, the error is found to be maximally $\pm 2\%$, with the exception of the first two years of the time series. Presumably, women could benefit from what had be learned about optimal training for the men's marathon in the previous decades, and hence could realize better potential in their early years of marathon racing. An important factor to consider when comparing track WRs to the half and full Marathon is that the latter are road races with much less controlled conditions like wind, exactly run distance, and elevation changes.

\section{Discussion}

\subsection{Interpretation of Changes in Model Parameters}

Since 1918, middle and long distance running has seen many changes. The aim of this section is to discuss some key observations that can explain the parameter improvements over a century. Some changes lead to major improvements, such as progress in sport science and coaching, new technologies, cultural aspects, while others had negative impacts. Exceptionally both gifted and hard-working runners and visionary coaches also participated in this progression.

\subsubsection{From generalist distance runners to specialized runners}

In the 1920th, the Finnish Paavo Nurmi broke 22 WRs between 1921 and 1930, from 1.500m to 20.000m. With such a dominance, Nurmi’s performances and his endurance is reflected in the parameter $E_\text{l}$, with a first peak at the beginning of the 1920th. The model parameters obtained from Nurmi’s personal best times are $t_c=10.8$min, $v_m=350.9$m/min, $d_c=3790$m, $E_\text{s}=0.36$, $E_\text{l}=6.87$ with a model error of 0.44$\%$. 
He was one of the first to combine both high-volume and high-intensity training.
In 1932, the French runner Jules Ladoumègue, the 1.500m world record holder at that time, was banned for life for professionalism by the International Federation IAAF \footnote{Athletisme magazine, \url{http://cdm.athle.com/asp.net/espaces.html/html.aspx?id=9634}}. The IAAF formally waved this professionalism prohibition in 1986, even if some more money was already present in the 1970th \citep{Henning:2020aa}. Those restrictions, with also training systems less developed, can explain that we see generalist runners, able to win medals and break record over a wide range of distances. The development of the professionalism in the 1970th lead to a higher interest in middle distance and distance running. This increased interest is reflected in the density of the Olympic Games results.
In 1939, the German Rudolph Harbig broke the 800m WR by a stunning 1.8sec. Harbig was also  very fast on shorter distances, as he broke the 400m WR in the same year, in 46.0sec. His training was based substantially on high-intensity training. Harbig’s 800m WR stood for an impressive 16 years period. Harbig’s career declined only during World War II.
In the 1940s, the Swedish Gunter Hägg broke 10 WRs, from 1.500m to 5.000m, and became the first man to run a sub-14 minute 5.000m, in 1942. At the opposite of Nurmi, Hägg’s dominance on middle distances lowered the endurance parameter $E_\text{l}$ to about $4$. In fact, this is confirmed by our model with parameters obtained from Hägg's personal best times which are $t_c=8.7$min, $v_m=369.8$m/min, $d_c=3217$m, $E_\text{s}=0.37$, $E_\text{l}=4.36$ with a model error of 0.42$\%$. 
In the 1950s, Emil Zatopek broke 18 WRs from 5.000m to 30.000m. He was able to sustain a very high training volume, up to 250km per week, something never seen before. A similar volume is logged by the current WR holder Kelvin Kiptum during the 
lead-up to a marathon, suggesting that this volume is probably close to optimal for elite marathon runners. At the 1952 Olympic Games, Zatopek won 3 gold medals, on the 5.000m, the 10.000m and the marathon, showing a rare exceptional endurance. Our model parameters for his personal best times (for distances from 1.500m to the marathon) suggest that he could sustain 90$\%$ of MAS (at a relative slow speed $v_m=345.1$m/min) for an enormous time $T_{90\%}=$87min. Indeed, Zatopek's performances helped to increase the endurance parameter $E_\text{l}$ in 1950, the year he improved his own WR over 10.000m by 18.6sec.  After Roger Banister achieved the first sub 4 Minute Mile in 1954, helped by 3 pacemakers, the pacemaker usage increased in the following decades, helping to break new WRs.

\subsubsection{A new area for athletics}

Runners from East Africa started to show to the world their talent in the 1960th, with the victory of the Ethiopian Abebe Bikila on the marathon at the 1960 Rome Olympic Games, running barefoot. The first Kenyan gold medal is won at the 1968 Mexico Olympic Games, on the 10.000m by Naftali Temu. They were the pioneers of a generation of African runners who would dominate middle distance running and distance running in the 1990th.
At the 1968 Mexico Olympic Games, synthetic track replaced for the first time clay tracks at the international level. This new surface material is known to return more energy and it is less sensible to rainy conditions. At the same Olympics, first doping tests were conducted. The first track runner to have failed a doping test at an Olympic Games is the Finnish Martti Vainio at the 1984 Olympic Games, after finishing 2nd at the 10.000m event.
In 1989, the first synthetic EPO was approved by the FDA. It turned out that EPO can profoundly increase maximal oxygen uptake (VO2max) and in fact it was being used to enhance athletic performance by the early 1990th. Improved techniques to detect use of exogenous EPO in 2005 limited ability to manipulate oxygen uptake and transport. This could be a plausible explanation for the saturation of WRs for men, and their MAS $v_m$. Other interpretation is that the biological or physiological limit of humans has been approached closely, after having optimized almost every (known) factor relevant for performance.

In the 1990th, women WRs dramatically improved, particularly thanks to Chinese athletes. In 1993, Wang Junxia broke the 10.000m WR by 41.96 sec, from 30:13.74 to 29:31.78, the by far largest improvement of a WR for that distance ever. Four days later, Qu Yunxia broke the 1.500m WR by 1.99 sec, in 3:50.46. And the day after, again Wang Junxia broke the 3000m WR in 8:06.11, a WR which is still standing 30 years later. All three WRs were established in Beijing, and the validity of those WR were questioned by the athletics observers. \footnote{In 2016, a letter, supposedly written by Wang Junxia, emerged accusing her coach Ma Junren that he forced her and her teammates to take “large doses of illegal drugs over the years” [\url{https://www.si.com/more-sports/2016/02/04/track-and-field-doping-china-wang-junxia-ma-army-letter}]} The IAAF (now World Athletics) opened an investigation, but no sanctions were taken.
In 2019, the World Anti-Doping Agency (WADA) reported 34.576 anti doping tests for Athletics, and these tests showed 173 anti-doping rule violations (0.5$\%$), including 103 for middle and long distance running \footnote{See \url{https://www.wada-ama.org/sites/default/files/2022-01/2019_adrv_report_external_final_12_december_2021_0_0.pdf}}. This shows that doping poses still a problem in running, and efforts should be made to detect suspicious athletes. As our model determines overall performance parameters from WR on various distances, a suspiciously fast WR can be easily noted by a large model error. Indeed, as can be seen from Fig.~\ref{fig:4}(a), the overall model error peaked in 1993 at 1.5$\%$, mainly driven by a 2.5$\%$ model error for the 3.000m, and a 2.0$\%$ model error for 1.500m. At the same time, the MAS $v_m$ showed a sudden jump, see Fig.~\ref{fig:4}(c). In the following years, the overall model error slowly decreased to about 0.6$\%$ nowadays, a value comparable to the one in the years before 1993. This indicates that it took about three decades for WRs on other distances to improve to a  level comparable to the 3.000m WR from 1993. Over this period of time the value of MAS changed only slightly, but showed a more marked increase in the last three years (see Sec.~\ref{sec:tech}).

\subsubsection{A new focus on the marathon}

For the marathon, nutrition, with carbohydrate loading and hydration strategies, has been improved in the 2000th. Also road running has been more attractive since the the 1990th with increased prize money, and many major cities starting to have their own marathon in the 1980th. Also, in the athletics world one was convinced that a runner should start racing on track distances and then move to the marathon later. An example for this is the runner Eliud Kipchoge, who was World Champion on the 5.000m in 2003, before starting his marathon career 10 years later. This belief was no longer true in the following years, after having seen the Kenyan Samuel Wanjiru winning the 2008 Olympic Games marathon at the age of 21. As a recent extreme, in 2023, Kelvin Kiptum broke the marathon world record at the age of only 23 years. Another important factor is the development of new racing shoes with a carbon fiber plate, released in 2017.
They would help reducing significantly the energetic cost of running for elite runners \citep{Hoogkamer:2018vg}.
 Indeed, Eliud Kipchoge using these new shoes broke the marathon WR in 2018 by 78 seconds, something never seen since the men's marathon  WR by Derek Clayton in 1967. Interestingly, we do observe an increase in endurance $E_\text{l}$ in 2020 due to new men WRs on 5.000m and 10.000m, both achieved with shoes containing PEBAX plates. 

For women, the British runner Paula Radcliffe revolutionized the marathon. In October 2001, the Kenyan Catherine Ndereba ran a marathon WR in 2:18:47. One year later, Radcliffe set a new WR in 2:17:18 and 6 months later she realized a stunning time of 2:15:25. This WR stood for 16 years, and was only broken by Brigid Kosgei (2:14:04) in 2019. Paula Radcliffe was a silver medalist on the 10.000m at the 1999 World Championships, and she made the difference on the marathon thanks to a very high endurance for which our model yields $E_\text{l} = 11.3$, based on her personal best times for distances from 1.500m to the marathon. In 2023, Ethiopia's Tigst Assefa became the first woman to run the marathon under 2 hours and 14 minutes, with a time of 2:11:53. This was a complete surprise as she had more modest  personal bests in the year before: 30:52 over 10km and 67:28 for the half marathon. Interestingly, for these two performances our model yields an endurance of $E_\text{l}=11.3$ which is identical to the one of Radcliffe. However, even with this excellent endurance a current world-class aerobic power with a MAS of about 370m/min is necessary for her WR, suggesting that Assefa should be able to run now much faster on 10km and half marathon. An endurance of about 11 was found to be maximal possible value as a function of training volume and intensity, by applying our model to data of thousands of marathon runners \citep{Emig:2020wj}. Both Kosgei and Assefa wore new technology shoes when achieving their records.
Hence it is instructive to compare the MAS of Radcliffe ($v_m=360$m/min from her personal best), achieved before the era of high-tech shoes, to the one estimated above for Assefa. The difference is 2.8$\%$ which is perfectly consistent with the measured improvement in running economy from fiber plate shoes in a WR holder \citep{Muniz_Pardos_2021}.

\subsubsection{Technology advancements since 2019}
\label{sec:tech}

 On the running track, athletes have been using new advanced shoe technology for spikes since 2019, in particular at the World Championships in Doha. This new technology would also improve running economy as fiber plate shoes do on the road \citep{Willwacher:2023aa}. A second new technology appeared in 2020 with the Wavelight pacing system.  It consists of light emitters along the inner curb of the running track. These emitters are programmed to provide a “moving light wave” at a freely programmable pace. For an analysis of the system, see \citep{Julin:2020aa}. Typically the wave speed is set to a new WR pace for world record breaking attempts. The pace needs not to be constant but can be programmed to change during a race. While pacing by technical devices in general was prohibited by World Athletics Technical Rules, in November 2019 a new rule was added which states  {\it
 "[...] the  following  shall  not  be  considered  assistance,  and 
are therefore allowed: Electronic  lights  or  similar  appliance  indicating  progressive times  during  a 
race, including of a relevant record."} \citep{Athletics:2023aa} 
 This system provides a visual advantage, and helps athletes to set a more even pace than it is possible by following pacemakers.
In 2020, Joshua Cheptegei improved the 5.000m WR by 1.99sec and the 10.000m WR by 6.53sec, following a very even pace strategy, guided by a Wavelight system. 
The previous WR were standing for respectively 16 years for the 5.000m and 15 years for the 10.000m, the longest period in history for both events. In fact, 400-meter splits of Cheptegei 10.000m race reveal an astonishingly even pattern with a standard deviation for the laps 2 to 24 of only 0.21sec. The previous WR on 10.000m was achieved by Bekele with a much larger standard deviation of 1.00sec for the same laps \citep{Julin:2020aa}.  Since 2020 almost every single new track WR for men and women has been realized with both new advanced technology spikes and the Wavelight system (with the exception of the women's 2.000m WR but including also Mo Farah's WR for the One Hour in 2020, and Lamecha Girma WR on 3.000m steeplechase in 2023).
For the women's WR on 5.000m in 2020, the Wavelight system was even set to an increasing pace requested by Letesenbet Gidey, allowing for a tailor-made, perfectly assisted negative split.  For women, the year 2023 was the one with most new WRs (1.500m, 1mile, 5.000m, marathon) since 1985. Due to recent new WR on the middle distances, the value of MAS increased after plateauing for almost three decades, see Fig.~\ref{fig:4}(c).
It can be concluded that those new technologies can explain the improvement of the endurance parameter $E_\text{l}$ for men and women since 2020, as the new spikes can help to reduce muscular fatigue and the Wavelight system mental fatigue as the athlete's mind can focus completely on the task of keeping up with the lights.

\subsection{Conclusions}

We have demonstrated that running WRs of more than a century can be described accurately by our mathematical model. The evolution of the model parameters over time could be explained in terms of
historical events such as changing training methodologies, social effects (professionalism, emergence of African runners, more women competing in races), technological advancements (shoe technology, pacing systems), and the use and control of performance enhancing drugs.
However, we note that a single new WR can have a substantial impact on model parameters which should be viewed as part of an adjustment process to a new overall performance level. This adjustment happens in a fluctuating manner as certain distances have bigger effects on a given parameter than other distances. For example, a change of the long distance endurance $E_\text{l}$ can arise from an exceptional WR on 3.000m (decreasing $E_\text{l}$), or on 10.000m (increasing $E_\text{l}$). We observe that over time the WRs on all studied distances have become increasingly mutually consistent which is reflected by an ever decreasing model error. 

Before, our model has been shown to describe the personal bests of individual runners \citep{Mulligan:2018ue,Emig:2020wj}. When looking at WRs, established by a group of runners, the model parameters reflect the overall “optimal” performance of a hypothetical generalist: In fact, marathon specialist Kipchoge, for example, might have a better endurance than the current WR parameters suggest, but his MAS might be a little slower than the value of $v_m$ of the WRs. But overall, it was beneficial for his performance on the marathon. We note that there are infinitely many combinations of model parameters that give the exact same marathon time. This flexibility of our model makes it applicable not only to individual runner's personal performances but also the best performances of a group of runners, i.e., WR performances. This is an important conclusion  for our approach. 

Our findings are different from the previously postulated power law relation between the mean running speed $v_\text{max}$ and distance $d$, $v_\text{max}\sim d^{-\beta}$ with an exponent $\beta$ that varies  between $0.054$ and $0.083$, depending on age and gender \citep{Riegel:1981ci}. Note that this exponent $\beta$ is smaller than the value $1/8$ expected from Kennelly's original work \citep{Kennely:1906aa}.  A modified, broken power law yielded a crossover duration $t_c$ between 3min and 4min which is too short to be consistent with direct laboratory measurements \citep{billat1998high}.

We have used our model to predict the best possible performance on the half and full marathon distance in each year, based on the WR performances on the track up to 10.000m in the same year. The accuracy of our prediction for both men and women, especially in recent years, appears surprisingly good, given the less controlled conditions in road races. In fact, when large efforts are made to simulate almost perfect racing conditions for a marathon, the finishing time becomes faster than what one would expect from track WRs. This is impressively demonstrated by the first sub-two hour marathon (time 1:59:40, Kipchoge, Vienna, 2019) which is not recognized as a WR due to artificial conditions. The course was perfectly flat with minimal curves and wind protection, an optimized configuration of seven pacemakers, guided by lasers, was employed to minimize air resistance over the entire distance, hydration was provided from a bicycle, and ideal weather conditions prevailed. 
It should be noted that even in the absence of wind, drafting can provide a metabolic power reduction of about $2\%$ \citep{Polidori:2020to}. In track races, important factors that reduce performance, relative to ideal conditions described above, are the relative large curvature of the oval track and the absence of air resistance reducing pacemakers over the entire distance.

Future studies based on our model could include the analysis of not only WRs but a given number of best performances per distance and year. This would provide information on the distribution of performance parameters for a given period of time. From this information the probability for new WRs could be estimated.

\section*{Conflict of Interest Statement}
The authors declare that the research was conducted in the absence of any commercial or financial relationships that could be construed as a potential conflict of interest.

\section*{Author Contributions}
TE: conceptualization, visualization, supervision, data anlysis, writing original manuscript. GA: data analysis, writing original manuscript.

\section*{Funding}
The author(s) declare no financial support was received for
the research, authorship, and/or publication of this article.

\section*{Data Availability Statement}
The datasets analyzed for this study can be found in the Supplemental Material.

\bibliographystyle{Frontiers-Vancouver} 



\section*{Figures}


\end{document}